\documentclass[preprint,12pt]{elsarticle}

\usepackage{amssymb}
\usepackage{amsmath}
\usepackage{enumitem,hyperref,dsfont}
\usepackage{amsthm}

\journal{}

\begin{document}
	
	\begin{frontmatter}
		
		
		
		\title{Contributions to the study of time dependent oscillators in Paul traps. Semiclassical approach.} 
		
		
		\author{Bogdan M. Mihalcea} 
		
		\ead{bogdan.mihalcea@inflpr.ro}
		\ead[url]{https://quest.inflpr.ro}
		
		\affiliation{organization={Natl. Inst. for Laser, Plasma and Radiation Physics (INFLPR)},
			addressline={Atomistilor Str. Nr. 409}, 
			city={Magurele},
			postcode={077125}, 
			state={Ilfov},
			country={Romania}}
		
		\begin{abstract}
		We investigate quantum dynamics for an ion confined within an oscillating quadrupole field, starting from two well known and elegant approaches. It is established that the Hamilton equations of motion, in both Schr\"{o}dinger and Heisenberg representations, are equivalent to the Hill equation. One searches for a linear independent solution associated to a harmonic oscillator (HO). An adiabatic invariant, which is also a constant of motion, is introduced based on the Heisenberg representation. Thus, the state of the non-autonomous system can be determined at any subsequent moment of time. The quantum states for trapped ions are demonstrated to be Fock (number) states, while the exact solutions of the Schr\"{o}dinger equation for a trapped ion are exactly the quasienergy states. Semiclassical dynamics is also investigated for many-body systems of trapped ions, where the wavefunction associated to the Schr\"{o}dinger equation is prepared as a Gauss package multiplied by a Hermite polynomial. We also discuss time evolution for the system under investigation and supply the propagator. 
		\end{abstract}
			
		\begin{highlights}
			\item Quantum states for trapped ions are demonstrated to be Fock (number) states, while the exact solutions of the Schr\"{o}dinger equation for a trapped ion are exactly the quasienergy states
			\item We investigate quantum dynamics (semiclassically) for a system of trapped ions and prepare the associated wavefunction as a Gauss package multipled by a Hermite polynomial. Time evolution of the system is discussed by means of the propagator which we supply
		\end{highlights}


\end{frontmatter}

\section{Introduction}\label{Intro}

Particle dynamics within a rapidly oscillating field has been firstly investigated by Kapitsa \cite{Kap51} based on a classical mechanics picture. More in-depth approaches have been used by Gaponov \cite{Gap58} or Landau and Lifshitz \cite{Land60}. A large amount of effort has been invested in investigating Hamilton functions with time dependent coefficients, where one of the most typical cases is the harmonic oscillator (HO) with time-dependent frequency and/or mass. Furthermore, the HO with time dependent frequency stands out as the first perfectly computed problem, by employing the Lewis-Riesenfeld (LR) invariant operator method \cite{Lew69, Mand17}. It relies on building an invariant operator and then expressing the Schr\"{o}dinger’s wave function in terms of invariant operator eigenstates with time dependent phase factor.
Furthermore, it was demonstrated that the dynamics of a particle confined within a high-frequency (RF) time-dependent potential is governed by a Schr\"{o}dinger equation characterized by a time-independent effective potential \cite{Cook85}, while the stability regions of classical and quantum-mechanical motion are identical. It was also emphasized that the prevailing effect of the time-dependent potential is to multiply the wave function of the static pseudopotential by a time-dependent phase factor. Then, Combescure investigated the evolution in time for Schr\"{o}dinger operators with time-periodic potentials when the classical equations of motion possess periodic orbit, while also demonstrating that various types of time-dependent Schr\"{o}dinger hamiltonians can be reduced to the quantum dynamics of the HO \cite{Comb86}. Quantum dynamics of single or many-body trapped ion systems is investigated extensively in \cite{Ste92, Schne12b, Fou19, Mih21}.

The LR approach has also been employed to characterize an ion confined in a Paul trap, treated as a quantum HO (QHO) \cite{Mih09, Kel21}, where state vectors have been defined based on a Fock state basis. An invariant operator can be associated with this system, which is exactly the quantum Hamiltonian. The basis states form a complete and orthogonal set of eigenvectors, which allows one to determine the spectrum of the quasienergy operator. Quantum motion of a particle in a nonlinear Paul trap has been investigated in \cite{Mih10a}, where the Hamilton equations of motion are inferred by applying the Time Dependent Variational Principle (TDVP) \cite{Kra81, Mih17}. It is demonstrated that the expected value of the Hamiltonian results as a function of the eigenvalues of the creation and annihilation operators. After using the TDVP and expressing the Hamilton equations of motion for a charged particle confined within a nonlinear Paul trap, the quantum equation of motion in the Husimi (Q) representation for the boson results that is fully consistent with the classical equation of motion of the perturbed oscillator \cite{Mih10a}. 

The TDVP method was also applied on coherent state orbits, which yields Hamilton type equations of motion on K\"{a}hler sub-manifolds such as classical phase spaces. Such formalism was employed to characterize Hamilton functions that are nonlinear in the infinitesimal generators of a dynamical symmetry group, such as nonlinear 3D ion traps \cite{Mih17, Mih18}. If the anharmonic part is a polynomial function, the Hamilton function describes an algebraic model that is linear in case of quadrupole ion traps (QIT). In agreement with the TDVP applied on coherent states, if an anharmonic analytical electric potential is added, then the classical equations of motion turn into coupled nonlinear Hill differential equations. Moreover, the quasienergy spectrum results from the solutions of the Schrödinger equations, expressed as the product between a geometrical (Berry) phase factor and the corresponding symplectic axial or radial coherent states \cite{Mih17}.

Most of the references mentioned above use the semiclassical approach, where the RF trapping field is not quantized but it is treated as a classical electromagnetic field.  Ref. \cite{Pedro21a} investigates quantum dynamics of an ion confined within a Paul trap, characterized by a time-dependent mass that increases exponentially and a trap frequency that falls exponentially in time. By making use of the LR invariant theory and Fock states, the time-dependent Schr\"{o}dinger equation for this problem is solved, whilst its solutions are employed to construct coherent states for the quantized particle in the Paul trap. Both Fock and coherent states are used to compute quantum observables, such as expectation values of the position and momentum operators, their variances and the uncertainty principle.

The paper brings new contributions towards investigating quantum dynamics of charged particles confined within an oscillating quadrupole field, such as the case of ions levitated in electrodynamic (Paul) traps \cite{Mih21, Mih23a}.  Applications span emerging quantum technologies (QT) and based on ultracold ions, including quantum sensors and new physics beyond the Standard Model (BSM) \cite{Schko22, Leib24}. 

The structure of the paper is discussed in the following. Section \ref{Intro} is the introductive part, which discusses the history and some of the achievements with respect to particle dynamics in rapidly oscillating fields. Section \ref{ConstMot} starts from the approaches used in Refs. \cite{Gla93, Leibf03} to discuss quantum dynamics of ions confined within an oscillating quadrupole field, such as the case of electrodynamic (Paul) traps. The section discusses techniques related to both the Schr\"{o}dinger and Heisenberg representations for a particle located within an oscillating field. Different solutions for the system under investigation are presented at the end of this section. Section \ref{EvolOp} reviews some of the results obtained in \cite{Mih22}, with respect to the time evolution of many-body systems of trapped ions confined in
a 3D quadrupole ion trap (QIT) that exhibits axial (cylindrical) symmetry. Section \ref{TrapDyn} brings original results and demonstrates that the wavefunction associated to the Schrödinger equation that describes a system of trapped ions with $n$ degrees of freedom can be engineered as a Gauss package multiplied by a Hermite polynomial. The propagator associated to this system is also derived.

\section{Constants of the motion. Glauber picture}\label{ConstMot}

Starting from the elegant approaches used in \cite{Gla93, Leibf03}, we investigate the quantum dynamics of a charged particle (ion) confined in an oscillating quadrupole field \cite{Pedro05, Onah23}, where the time-dependent wave functions are considered to exhibit a simple one-to-one correspondence with the wave functions associated with an ordinary static-field HO. The oscillating potential of a Paul trap is quadratic along both axial and radial components of the motion, which renders the dynamics separable. Moreover, the Kapitsa approximation \cite{Kap51} is valid both in the classical and quantum approach cases. When using the quantum approach, one assumes the time-dependent potential characteristic to an electrodynamic (Paul) trap  is quadratic in all Cartesian coordinates of the center of mass (CM) of the confined ion. The Hamilton function that characterizes the trapped ion is \cite{Gla93, Leibf03}

\begin{equation}\label{Gla1}
	H_0\left(q, p, t\right) = \frac{p^2}{2m} + \frac m2 \ W\left(t\right)q^2 \ ,
\end{equation} 
where $q$ and $p$ stand for the coordinate and momentum of the trapped ion, while 
\begin{equation}\label{Gla12}
	W\left(t\right) = \frac{\Omega^2}4\left(a + 2q \cos{\Omega t}\right)
\end{equation}
 depicts the time dependence of the applied external oscillating field of radiofrequency denoted by $\Omega$. The Hamilton equations of motion are \cite{Cook85, Gla93, Leibf03, Meni07}

\begin{equation}\label{Gla2}
	\begin{cases}
		\dot q = \frac pm \ , \\
		\dot p = - m W\left(t\right) q \ .
\end{cases}
\end{equation}
One can use the Heisenberg picture to write the equations of motion as follows \cite{Leibf03}

\begin{equation}\label{Gla21}
	\begin{cases}
		\hat{\dot q} = \frac 1{i\hbar}[\hat x, \hat H_0] = \frac {\hat p}m \ , \\
		\hat{\dot p} = \frac 1{i\hbar}[\hat p, \hat H_0] = - m W\left(t\right) \hat q \ .
	\end{cases}
\end{equation}
 where both eqs. (\ref{Gla2}) and (\ref{Gla21}) can be expressed as

\begin{equation}\label{Gla3}
	\ddot q + W\left(t\right)q = 0 \ .
\end{equation}
Eq. (\ref{Gla3}) can be considered as a Hill equation \cite{Bril48, Lach64, Kova18} by replacing $q$ with the complex c-number function $u\left(t\right)$. As explained in \cite{Mih22}, a c-number is a {\em classical} number and it is related to any number or quantity which is not a quantum operator, but acts upon elements of the Hilbert space of states of a quantum system. Hence, c-numbers represent ordinary complex-valued quantities whose algebra is commutative. Therefore, one can consider the operator formalism as a generalized version of the probability theory, in which real-valued random variables are represented by self-adjoint operators on a Hilbert space \cite{Whit00}. In addition, complex-valued random variables are represented by normal operators. We also consider $q_1\left(t\right)$ and $q_2\left(t\right)$ as two differentiable functions that are solutions of eq. (\ref{Gla3}). If a linear independent solution $q = c_1 q_1 + c_2q_2$ of the system exists, the Wronskian square matrix (determinant) should be nonsingular, case when

\begin{equation}\label{Gla4}
	q_1 \dot q_2 - \dot q_1 q_2 = const  \ ,
\end{equation}  
which yields a constant of motion for the system. If the Wronskian is zero, the solutions $q_1$ and $q_2$ are linearly dependent. Glauber \cite{Gla93} demonstrates that an alternative approach is to regard $q_1$ as an ordinary c-number solution \cite{Mih22} of eq. (\ref{Gla4}), and consider $q_2$ to be the general operator solution $q\left(t\right)$ to eqs. (\ref{Gla2}), regarded as Heisenberg equations of motion. Hence, one introduces the complex function $u\left(t\right)$ defined as \cite{Gla93, Leibf03}

\begin{equation}\label{Gla5}
	\ddot u + W\left(t\right) = 0 , \ \mbox{with initial conditions} \  u\left(0\right) = 1 , \ \ \dot u\left(0\right) = i\omega \, 
\end{equation} 	
where $\omega \in \mathbb R$ is a real parameter. What is more, the solution $u\left( t\right)$ also satisfies the Wronskian identity \cite{Leibf03} 

\begin{equation}\label{Gla6}
	W = \begin{vmatrix}
		u^* & u \\
		\dot u^* & \dot u
	\end{vmatrix} =
	\begin{vmatrix}
		1 & 1 \\
		-i\omega & i\omega
	\end{vmatrix} =
	2i\omega \ .
\end{equation}
One chooses a general solution expressed as

\begin{equation}
	u_{gen} = Au + Bu^* \ ,
\end{equation}
where $u^*$ denotes the complex conjugate of $u$. If

\begin{equation}
	u_{gen} = u^*_{gen} \Rightarrow A^* = B \ .
\end{equation}
Further on we choose $W > 0, W = \ \mbox{const}\ $ and $\omega = \sqrt{W}$. If one denotes 

\begin{equation}
	u = e^{i\omega t} \Rightarrow \dot u = i\omega u \ ,
\end{equation}
and 

\begin{equation}
	u_{gen} = A e^{i\omega t} + A^* e^{i\omega t} \ ,
\end{equation}
which describes the equation associated with a harmonic oscillator (HO). One can also write

\begin{equation}
	u_{gen} = M\cos{\omega t} + N\sin{\omega t} \ ,
\end{equation}
with

\begin{equation}
	\begin{cases}
		M = A + A^* = 2 \Re e A \ , \\
		N = i\left(A - A^*\right) = -2 \Im m A \ .
	\end{cases}	
\end{equation}
Hence, 

\begin{equation}
	u_{gen} = \sqrt{M^2 + N^2} \cos\left(\omega t + \varphi\right) \ ,
\end{equation}
with

\begin{equation}\label{Gla10}
	\begin{cases}
		M = \sqrt{M^2 + N^2} \cos\varphi , \\
		N = - \sqrt{M^2 + N^2} \sin \varphi \ .
	\end{cases}	
\end{equation}
We revert to eq. (\ref{Gla6}) and require the Wronskian determinant to be constant in time. One introduces the function  \cite{Gla93}

\begin{equation}\label{Gla11}
	C\left(t\right) = \frac 1{\sqrt{2m\hbar \omega}}\left[u\left(t\right) p\left(t\right) - m \dot u\left(t\right)q\left(t\right)\right] \ ,
\end{equation}  
which is an adiabatic invariant \cite{Hen93} and an operator constant of the motion \cite{Sim24}. By further deriving eq. (\ref{Gla11}) one infers

\begin{equation}\label{Gla12}
	\dot C\left(t\right) = \frac 1{\sqrt{2m\hbar \omega}}\left[\dot u p + \dot p u - m \ddot u q - m\dot u \dot p \right] \ .
\end{equation}
We use the Hamilton equations (\ref{Gla2}) and obtain

\begin{equation}\label{Gla13}
	\dot C\left(t\right) = \frac 1{\sqrt{2m\hbar \omega}}\left[\dot u p - m W\left(t\right) q u - m \ddot u q - \dot p u \right] = 0 \ .
\end{equation}
As an adiabatic invariant can only be quantified in the Heisenberg representation \cite{Sim24, Zele11}, $C\left(t\right) = C\left( 0 \right) = \ \mbox{const}$ and \cite{Gla93, Leibf03}

\begin{equation}\label{Gla14}
	C\left(0\right) = \frac 1{\sqrt{2m\hbar \omega}} \left( p - im\omega q\right) \ ,
\end{equation}  
stands for the classical constant of motion. Furthermore, in the quantum approach

\begin{equation}\label{Gla15}
	a = \frac 1{\sqrt{2m \hbar \omega}}\left( \hat p - im\omega \hat q\right) \ ,
\end{equation}
represents the non-Hermitian annihilation operator \cite{Gla93, Dodon02} with respect to excitations of a static-field oscillator of frequency $\omega$. Thus, by introducing the vacuum state vector $|\phi_0\rangle$ and the ground state vector $| 0 \rangle$, one can write 

\begin{equation}\label{Gla16}
	\begin{cases}
	a |\phi_0 \rangle = 0 \ , \\
	a |0 \rangle = 0 \ .
	\end{cases}
\end{equation}

By using the Heisenberg representation \cite{Gla93, Leibf03}, the constant of motion is expressed as

\begin{equation}\label{Gla17}
	\hat C = \frac i{\sqrt{2m\hbar \omega}} \left( u\hat p - m\dot u \hat q\right) \ ,
	\hat C |0\rangle> = 0 \ ,
\end{equation}  
Thus, one can determine the state of the non-autonomous system at any subsequent moment of time.
The operator $C\left(t\right)$ in eq. (\ref{Gla12}) is defined within the Heisenberg picture as a linear combination of the operators $q\left(t\right)$ and $p\left(t\right)$. In addition, the unitary transformation from the Schr\"{o}dinger to the Heisenberg picture is defined as

\begin{equation}\label{Gla18}
	\hat q = U^{-1}\left(t\right) q U\left(t\right) \ , \ U\left(0\right) = 1 \ ,
\end{equation}
where $U$ denotes a time-dependent unitary operator of initial value $U\left(0\right)$, whilst $\hat q$ denotes the Heisenberg representation and $q$ denotes the Schr\"{o}dinger representation. In addition 

\begin{equation}
	q = \hat q\left(0\right) \ .
\end{equation}
It is not necessary to find $U\left(t\right)$. We just observe that $\hat C\left(t\right)$ is related to its Schrodinger-picture counterpart $C_{Sch}\left(t\right)$ by the relation

\begin{equation}\label{Gla19}
	\hat C = U^{-1}C_{Sch} U \ ,
\end{equation}
where 

\begin{equation}\label{Gla20}
	C_{Sch}\left(t\right) = \frac i{\sqrt{2m\hbar \omega}}\left[u\left(t\right) p - m\dot u\left(t\right)q\right] \ .
\end{equation}
Besides, one also writes

\begin{equation}\label{Gla201}
	C_{Sch}\left(t\right) U\left(t\right) |0 \rangle = 0 \ ,
\end{equation}
with

\begin{equation}\label{Gla22}
	U\left(t\right) |0\rangle = |0, t\rangle \ , \ \mbox{and} \ U\left(t\right)|n\rangle = |n, t\rangle \ ,
\end{equation}
where $|n\rangle$ stands for the oscillator states (also denoted as number or Fock states). As demonstrated in \cite{Leibf03}, there is a close correspondence between confinement of ions within a Paul trap or within a static potential. Therefore, it is useful to express the states of motion for trapped ion systems based on the oscillator number operator. Using the Heisenberg picture \cite{Zele11, Sim24} one can write

\begin{equation}\label{Gla23}
	\hat N = \hat C^{\dagger}\left(t\right) \hat C\left(t\right) = \hat a^{\dagger} \hat a \ .
\end{equation}  
As $\hat c\left(t\right)$ is time independent (see eq. \ref{Gla13}), it is evident that the $\hat N$ operator is time-independent and its eigenstates are exactly the Fock or number states \cite{Mih09, Dodon02} with the corresponding ladder algebra

\begin{equation}\label{Gla24}
	\begin{cases}
	\hat a| n\rangle = \sqrt n\left(n - 1\right) \ , \\
	\hat a^{\dagger}| n \rangle = \sqrt {n + 1}\left(n + 1\right) \ , \\ 
	\hat N |n \rangle = n |n \rangle \ ,
	\end{cases}
\end{equation}
where $a^{\dagger}$ is the creation non-Hermitian operator while $a$ stands for the annihilation operator. By turning to the Schr\"{o}dinger picture and using eq. (\ref{Gla22}) one derives

\begin{equation}\label{Gla25}
	\hat U^{\dagger}\left(t\right) \hat N \hat U\left(t\right) = \hat U^{\dagger}\left(t\right) \hat C^{\dagger}\left(t\right) U\left(t\right) \hat U^{\dagger}\left(t\right) \hat C\left(t\right) U\left(t\right) = \hat C_{Sch}^{\dagger} \left(t\right)\hat C_{Sch} \left(t\right) \ .
\end{equation}
The eigenstates and eigenvalues of these operators are \cite{Leibf03}

\begin{equation}\label{Gla26}
	\begin{array}{ll}
		\hat C_{Sch}\left(t\right)|n, t> = \sqrt{n}|n - 1, t> \ , \\
		\hat C_{Sch}^{\dagger} \left(t\right)|n, t> = \sqrt{n + 1}|n + 1, t> \ ,
	\end{array}
\end{equation}
which results in

\begin{equation}\label{Gla27}
	\hat N_{Sch}\left(t\right) |n, t> = n|n, t> \ .
\end{equation}

Hence, it is demonstrated that the Schr\"{o}dinger-picture eigenstates can be employed in complete analogy to the static potential HO, whilst all algebraic properties of the static potential ladder operators carry over to $\hat C_{Sch}\left(t\right)$ and $\hat C_{Sch}^{\dagger}\left(t\right)$. Owing to the micromotion which periodically alters the ion kinetic energy, it stands to reason to link the quantum number $n$ with the energy of the ion averaged over a period of the RF drive frequency. This is exactly the case of the pseudopotential approximation \cite{Deh68, Robe18}, when the electric forces that are time-averaged over an RF period generate a harmonic potential \cite{Kno15, Kaji22}).    

\begin{equation}\label{Gla28}
	C_{Sch}\left(t\right) |0, t\rangle = 0 \ ,
\end{equation}
 where
 
 \begin{equation}\label{Gla29}
 	\phi_n\left(q, t\right) = \langle q | n, t\rangle \ ,
 \end{equation}
with

\begin{equation}\label{Gla30}
	\left[ -i\hbar u \frac \partial{\partial q} - m\dot u q\right] \phi_0\left(q, t\right) = 0 \ .
\end{equation}
The normalized solution of eq. (\ref{Gla30}) is expressed as \cite{Gla93}

\begin{equation}\label{Gla31}
	\phi_0 \left(q, t\right) = \left(\frac {m\omega}{\pi \hbar}\right)^{1/4}\frac 1{u\left( t \right)^{1/2}} \exp \left( \frac {im}{2\hbar} \frac{\dot u\left(t\right)}{u\left(t\right)}q^2\right) \ ,
\end{equation}
which describes a squeezed coherent state, while it also stands as the wavefunction associated to an ion confined in a Paul (RF) trap. This is also a Gaussian wave function quite similar to the ground state wave function for the static-field oscillator. The excited states are described by

\begin{equation}\label{gla32}
	\phi_n\left(q, t\right) = \frac 1{\sqrt{n!}}\left(\frac{u^*}{2u}\right)^{n/2}H_n\left[\left(\frac{m\omega}{\pi \hbar |u\left(t\right)|^2} \right)^{1/2} q\right] \cdot \phi_0\left(q, t\right) \ .
\end{equation}
The functions above are also called quasienergy functions, while $H_n$ stands for the Hermite polynomial of order $n$. Thus, the exact solutions of the Schr\"{o}dinger equation for a charged particle in a Paul trap are exactly the quasienergy states \cite{Mih18, Mih22, Ghe92}. The expectation values of $q^2\left(t\right)$ and $p^2\left(t\right)$ are discussed in \cite{Gla93}. 

The classical micromotion appears in the wavefunction as a pulsation at the period of the driving field \cite{Leibf03}. When $W = \omega_p^2, \ \omega_p > 0$, where $\omega_p$ is the frequency associated to the pseudopotential \cite{Paul90, Kno15, Kaji22}, one writes

\begin{equation}\label{gla33}
	\phi_{pn}\left(q, t\right) = \frac{\left(i\omega_p\right)^{n/2}}{\sqrt{n! \cdot 2^n}} H_n\left[\sqrt{\frac{m\omega_p}\hbar}\cdot q\right] \cdot \phi_{p0}\left(q, t\right) \ ,
\end{equation}
and

\begin{equation}\label{gla34}
	\phi_{p0}\left(q, t)\right) = \left(\frac{m\Omega}{\hbar \pi}\right)^{1/4} \frac 1{u\left(t\right)} \cdot \exp{\left(\frac{-m\Omega}{2\hbar}\ q^2\right)} \ .
\end{equation}
Eq. (\ref{gla34}) describes a HO (trapped ion) of pulsation (frequency) $\omega_p$, whilst the mid term $1/u\left(t\right)$ is unitary. We also denote \cite{Mih09}

\begin{equation}\label{gla35}
	f_n = \langle\phi_{pn} | \phi_n \rangle \ .
\end{equation}

For a static potential HO, the evolution of the energy eigenstates multiplies the wave function by a phase factor (which is why they are called stationary states) \cite{Ghe92, Leibf03, Mih17}. In case of a time-dependent potential (non-autonomous Hamilton function), {\em e.g.} for an ion confined in a Paul trap, the statement above is valid, but only for integer multiples of the RF period $T=2\pi/\Omega$. Thus, the corresponding states given by eq. (\ref{gla33}) are not energy eigenstates (because of the periodic energy transfer from the RF field, in analogy to the classical micromotion), but they can be considered as the closest approximation to stationary states within a time-dependent potential. Owing to this, they are called quasistationary states. 

The full quantum treatment is available in \cite{Leibf03}. In addition, recent quantum engineering techniques enable one to implement controllable spin-spin interactions between ions, making trapped ions an almost ideal quantum simulator platform \cite{Gras14, Berm17, Katz23}. 

\section{Evolution operators for systems of trapped ions}\label{EvolOp}

In the following we will remind some of the results in Ref. \cite{Mih22}, with respect to the time evolution of a system consisting of $N$ identical ions of mass $m$, confined in a 3D quadrupole ion trap (QIT) that exhibits axial (cylindrical) symmetry \cite{Mih17}. The trap is characterized by a constant axial magnetic field and a time varying quadrupole electric potential, which is the standard case of a combined (Paul and Penning) trap, where the residual interaction can be treated as a perturbation. Both axial $\left( n = N\right)$ and radial $\left( n = 2N\right) $ dynamics for a system of trapped ions which exhibit $n$ degrees of freedom, is characterized by the quantum Hamilton function

\begin{equation}\label{ber1}
	H = \sum_{k = 1}^n \left( - \frac{\hbar^2}{2m} \frac{\partial^2}{\partial x_k^2} + \lambda \frac m2 x_k^2 \right) + V \left( \vec {\textbf x} \right) + c \left( \vec {\textbf x} \cdot {\vec {\textbf p}} + \vec {\textbf p} \cdot {\vec {\textbf x}} \right) \ .
\end{equation}

One denotes $\vec{\textbf x} = \left(x_1, x_2, \ldots, x_n\right)$ and $\vec{\textbf p} = \left(p_1, p_2, \ldots, p_n\right)$, where $p_k = - i \hbar \partial /\partial x_k$ represents the quantum momentum operator. Furthermore, it is assumed the control parameters $\lambda $ and $c$ are periodic in time, whilst $V$ stands for a homogeneous potential, invariant with respect to translations of order $-2$ . The issue of the physical interpretation of these control parameters for Paul, Penning and combined traps is explicitly treated in Refs. \cite{Mih17, Major05}.  

Consequently, one writes the solution of the time dependent Schr\"{o}dinger equation as \cite{Comb86}

\begin{equation}\label{ber2}
	\Phi \left( t\right) = U\left(t, t_0\right) \Phi \left( t_0\right) \ ,
\end{equation}
where $U\left(t,  t_0\right) $ denotes the unitary evolution operator that can be explicitly built as  

\begin{equation}\label{ber3} 
	U\left( t, t_0\right) = S\left( t\right) \exp \left[- i\left( \tau - \tau_0\right) H_0\right] S^{-1}\left( t_0\right)\ ,
\end{equation} 
with

\begin{subequations}\label{ber4}
	\begin{eqnarray}
		S\left( t\right) = \exp \left( i\alpha \vec{\textbf z}\ ^2/2\right) \exp \left[ -i\beta \left( \vec{\textbf z}\cdot {\vec{\boldsymbol \nabla }} + {\vec {\boldsymbol \nabla }} \cdot \vec{\textbf z} \right) \right] \ , \\
		H_0 = - \frac 12 {\vec{\boldsymbol \nabla}^2} + m \hbar^{-2} V\left( \sqrt{\hbar /m}\  \vec{\textbf z}\right) + \frac 12 \vec{\textbf z}^2\ , \\
		\vec{\textbf z} = \sqrt{m/\hbar }\ e^{- \alpha }\ \vec{\textbf x}\;,\ \vec{\boldsymbol \nabla} = \left(\partial/\partial z_1, \ldots, \partial /\partial z_n\right) \;.	
	\end{eqnarray}
\end{subequations}

The time dependent functions $\alpha$, $\beta$ and $\tau $ result out of the Hill equation \cite{Bril48, Lach64, Kova18}
\begin{equation}\label{ber5}
	\ddot \zeta + \left(\lambda - 2 \dot c - 4 c^2\right) \zeta = 0 \ ,
\end{equation}
with $\zeta = \exp \left( \alpha + i \tau \right),\; 2\beta = \dot \alpha - 4c$. The case of a 3D QIT $\left(c = 0\ ,\; V = 0\right)$ is investigated in \cite{Comb86}. The stability regions characteristic to this equation determine the control parameters for stable classic motion. Quantum stability in a 3D ideal QIT is characterized by a discrete quasienergy spectrum \cite{Mih18, Ghe92, Major05}
$$
\varepsilon_j= \mu \left(2j + E_0\right) ,\;\;j = 0, 1, \ldots ,
$$
where $\mu E_0$ is the fundamental state quasienergy, and $\mu$ stands for the Floquet exponent that corresponds to the classical stability regions \cite{Bril48, Lach64, Kova18, Mih24a}. As demonstrated in \cite{Ghe92, Mih17, Mih18}, the quantum quasienergy states are symplectic coherent states (CS). One employs the analytical model introduced in \cite{Mih21}, which uses the relative coordinates $y_{\alpha j}$ along with a set of collective variables $s$:

\begin{equation}\label{ber6}
	y_{\alpha j} = x_{\alpha j} - \frac 1N \sum_{\alpha = 1}^N x_{\alpha j,} \ ,\;s = \sum_{j = 1}^d \sum_{\alpha = 1}^N y_{\alpha j}^2 \ ,
\end{equation}
with $x_{\alpha j} = x_{\alpha + j\left(N - 1\right) },\;1\leq \alpha \leq N,\;1\leq j\leq d$. By choosing an electric potential such as

\begin{equation}\label{ber7}
	W = \frac{bs}2 + 2a\sum_{\mu ,\nu }C_{\mu \nu }V_{\mu \nu }\ , \;
	V_{\mu \nu } = s^{-\mu }\sum_{\alpha \neq \beta }\left( \sum_{j = 1}^d \left(x_{\alpha j} - x_{\beta j}\right)^2 \right)^{\nu - 1} \ .
\end{equation}

As demonstrated in \cite{Mih22}, integrable Hamilton functions characterized by a discrete quasienergy spectrum can be obtained if $C_{\mu \nu } = 0$ for $\mu = \nu$. The Coulomb interaction is defined by $\mu = 0$ and $\nu = 1/2$. The classical Hamilton function is then determined by the expectation values of the quantum Hamilton function $H$ on coherent oscillator and symplectic states \cite{Ghe92, Mih17}. The equilibrium points are derived as solutions of a system of $n = Nd$ equations:
\begin{multline}\label{ber8}
	by_{\alpha j} - a\sum_{\mu ,\nu }\mu s^{-1}y_{\alpha j}C_{\mu \nu }V_{\mu \nu}  \\
	+ 2 a\sum_{\mu, \nu}\left( \nu - 1\right) s^{-\mu }C_{\mu \nu }\sum_{\alpha \neq \beta }\left( x_{\alpha j} - x_{\beta j}\right) \left( \sum_{k = 1}^d \left(x_{\beta k} - x_{\alpha k}\right)^2\right)^{\nu -2} = 0 \ .
\end{multline}

Ion crystals emerge in regions characterized by configurations of minimum \cite{Major05, Mih18, Mih21}. In case of a one dimensional (1D) integrable dynamical system consisting of $N$ particles, that corresponds to the Calogero potential \cite{Calo71} 

\begin{equation}\label{ber9}
	V = g\sum_{\alpha \neq \beta }\left( \xi _\alpha -\xi _\beta \right)^2, \quad \xi_\alpha = \left( b/ag\right)^{-4} x_{\alpha 1}\ , g > \frac {-\hbar^2}{4m} \ ,
\end{equation}
where the coordinates of the equilibrium configuration $\xi_\alpha $ are exactly the zeros of the Hermite polynomial of order $N$. These solutions completely determine the minimum points of the potential function, and consequently the ordered structures of ion crystals (strongly coupled Coulomb non-neutral plasmas) \cite{Leib24, Affo20, Stopp22}. 

\section{Trapped ion dynamics. Semiclassical approach.}\label{TrapDyn}

The Schr\"odinger equation that describes a system of trapped ions with $n$ degrees of freedom can be cast as 
 
\begin{equation}\label{mis1}
	i\hbar \frac{d\psi}{dt} = -\frac{\hbar^2}{2m} \bigtriangleup \psi
	+ V\left( t,{\mathbf x}\right) \ \psi \ ,\;\ 
	\bigtriangleup = \sum\limits_{i=1}^n\frac{\partial^2}{\partial x_i^2}\;,
\end{equation}
where ${\mathbf x} = \left( x_1,\ldots,x_n\right) $. If the potential $V$ is a polynomial of rank no higher than 2 in the coordinates ({\em e.g.}, in case of the $n$-dimensional parametric HO), then the solution of the Schr\"odinger equation (\ref{mis1}) can be expressed by means of the wavefunction $\Phi_k$ engineered as a Gauss package multiplied by a Hermite polynomial: 

\begin{multline}\label{mis2}
	\Phi_k\left( {\mathbf {A,B,q,p,x}}\right) =\frac{2^{-\left| k\right| /2}\left(	\pi \hbar \right)^{-n/4}}{\sqrt{k!\det A}}H_k\left( A;\hbar ^{-1/2}\left|A\right|^{-1}\left(\mathbf x-q\right) \right)\\
	\times \exp \left\{-\frac 1{2\hbar}\left\langle \left( {\mathbf {x-q}}\right), BA^{-1}\left( {\mathbf {x-q}}\right) \right\rangle + \frac i\hbar
	\left\langle {\mathbf p},\,\left( {\mathbf{x-q}}\right) \right\rangle \right\} \ .
\end{multline}

One denotes
\begin{equation}\label{mis3}
	\left\langle {\mathbf x,y}\right\rangle \equiv {\mathbf x}^\dagger{\mathbf y}\ ,\;\left\langle {\mathbf x}A{\mathbf y}\right\rangle = x^\dagger Ay\ ,
\end{equation}
where $\mathbf{x,y}\in \mathbb C^n$ are column vectors, $\mathbf x^\dagger$ stands for the Hermitian conjugate of the vector $\mathbf x$, whilst $A$ are $B$ are $n\times n$ dimension invertible, complex matrices, that satisfy

$$
BA^{-1} = \left( BA^{-1}\right)^t\;,\ \;A^\dagger B + B^\dagger A = 2I_n \ , 
$$
where $^t$ stands for the transpose of a matrix and $I_n$ denotes the $n\times n$ dimension unitary matrix. By employing the polar decomposition \cite{High86}, one finds
$A = U\left| A\right|$, where $U$ represents a unitary operator, while $\left| A\right| 
= \sqrt{A^{\dagger }A}$ is a positively defined Hermitian (symmetric) matrix, in such a way that all its eigenvalues are positive. 

We introduce the Hermite polynomials with multiple variables 

\begin{equation}\label{mis5}
	\tilde {H_0}\left( {\mathbf x}\right) = 1 \ ,\;\;\tilde {H_1}\left( {\mathbf {v_1; x}}\right) = 2\left\langle {\mathbf {v_1, x}}\right\rangle \ .
\end{equation}

We now define recursively the Hermite polynomial of rank $m$, denoted as $H_m$ 

\begin{multline}\label{mis6}
\tilde {H}_m\left(\mathbf v_1, \ldots, \mathbf v_m; \mathbf x\right) = 2 \left\langle
\mathbf v_m, \mathbf x\right\rangle \tilde {H}_{m-1}\left(\mathbf v_1,\ldots , 
\mathbf v_{m-1}; \mathbf x\right) \\
- 2 \sum_{i=1}^{m-1}\left\langle \mathbf v_m,{\mathbf v_i} \right\rangle \times \tilde H_{m-2}\left( \mathbf v_1,\ldots ,\mathbf v_{i-1}, \mathbf v_{i+1},\ldots , \mathbf v_{m-1};\,\mathbf x \right) \ ,
\end{multline}

\begin{equation}
	\label{mis7}H_k\left(A; \mathbf x\right) = \tilde H_{\left| k\right| }\left(
	\mathbf v_1, \ldots , \mathbf v_{\left| k\right| }\,; \mathbf x\right) \ ,\;\
	\left| k\right| = k_1 + \ldots + k_n \ ,
\end{equation}

\begin{equation}
	\label{mis8}\mathbf v_s = A\left( A^\dagger A\right)^{-1/2}\mathbf u_s\ ,\;\
	u_{si} = \delta_{ij}\ , \,k_{j-1} < s\leq k_j\ ,
\end{equation}
where $\delta_{ij}$ stands for the Kronecker delta function. Then, we introduce the classical trajectories for an ion of mass $m$ confined within a force field $-\nabla V\left( t, \mathbf q\right) $ (Paul trap):  

\begin{equation}\label{mis9}
	\dot q\left( t\right) = \frac 1m\, p\left( t\right) \ ,\;\ \dot p\left( t\right) = - \nabla V\left( t, q\left( t\right) \right) \ ,\;q\left(
	0\right) = q_0 \ ,
\end{equation}
with initial conditions $q\left( 0\right) = q_0$ and $p\left( 0\right) = p_0$. The classical action \cite{Land60, Mih17} is

\begin{equation}\label{mis11}
	S\left( t\right) = \int\limits_0^t\left[ \frac 1{2m}\ p^2\left(
	s\right) - V\left( s,q\left( s\right) \right) \right] ds\ .
\end{equation}
Further on, one introduces the complex matrices $A\left( t\right) $ and $B\left( t\right) $ as solutions of the classical equations of motion which satisfy

\begin{equation}\label{mis12}
	\dot A\left( t\right) =\frac im\ B\left( t\right) \ ,\;\
	A\left( 0\right) = A_0\ ,
\end{equation}

\begin{equation}\label{mis13}
	\dot B\left( t\right) = iV_H\left( t,q\left( t\right) \right)
	A\left( t\right) \ ,\;\ B\left( 0\right) = B_0\ ,
\end{equation}
where $V_H$ denotes the Hessian matrix of $V$ \cite{Mih21}. The propagator $U$ associated to the  Schr\"odinger eq. (\ref{mis1}) is characterized by 

\begin{equation}\label{mis14}
	U\left( t,t_0\right) \psi \left( t_0\right) =\psi \left(t\right) \ ,\;\ t_0\leq t\ ,
\end{equation}

\begin{equation}\label{mis15}
	U\left( t,\,t^{\prime }\right) U\left( t^{\prime },\,t"\right) 	= U\left( t,\,t"\right) \ ,\;\ 0\leq t\leq t^{\prime }\leq t"\leq T\ .
\end{equation}
Under these circumstances, for any $r\in \mathbb Z$, there is a uniform constant in $t$ and $\hbar $ denoted as $C$, in such a way that for $0\leq t\leq T$

\begin{equation}\label{}
\left\| U\left( t, 0\right) \psi \left( 0\right) - \exp \left( -i S\left(t\right) /\hbar \right) \psi \left( t\right) \right\| \leq C t\hbar ^{1/2}\ ,
\end{equation}
which enables one to determine the evolution of the system in time. 

\section{Discussion}\label{Disc}
Quantum dynamics of ions confined within an oscillating quadrupole potential (electrodynamic trap) is investigated, starting from the refined models presented in \cite{Gla93, Leibf03}. The Hamilton equations of motion, in both Schr\"{o}dinger and Heisenberg representations, can be regarded as a Hill equation, by introducing a complex c-number function. As shown in \cite{Mih22}, a c-number is a classical number which is related to any number or quantity which is not a quantum operator,
but acts upon elements of the Hilbert space of states of a quantum system. Therefore,
c-numbers are complex-valued quantities whose algebra is commutative. 

We search for linear independent solutions of the system, which requires the Wronskian matrix to be non-singular. It is demonstrated that the general solution is similar to the solution for the HO. Then, we introduce an adiabatic invariant which is also a constant of the motion, that we quantify in the Heisenberg representation. We show this operator can be used to determine the state of the system at any subsequent moment of time. The quantum states associated to the system are Fock (number) states or oscillator states, which proves the existing correspondence between confinement of ions within a Paul trap or within a static potential. Furthermore, the Schr\"{o}dinger picture eigenstates can be employed in complete analogy to the static potential HO. Because the micromotion periodically alters the ion kinetic energy due to the energy transfer from the driving field, it stands to reason to link the quantum number $n$ with the energy of the ion averaged over a period of the RF drive frequency. This is exactly the case of the pseudopotential approximation \cite{Deh68, Major05}, when the electric forces that are time-averaged over an RF period generate a harmonic potential \cite{Kaji22}. The normalized solution describes a squeezed coherent state. Therefore, the exact solutions of the Schrödinger equation for a charged particle in a Paul trap are exactly the quasienergy states.

We finally investigate the Schrödinger equation that describes a system of trapped ions with $n$ degrees of freedom. If the potential $V$ is a polynomial of rank no higher than 2 in
the coordinates (e.g., in case of the n-dimensional parametric HO), we demonstrate the solution
of the Schr\"{o}dinger equation can be expressed as a wavefunction that is engineered as a Gauss package multiplied by a Hermite polynomial. We also derive the propagator associated to the system.

\section{Acknowledgements}
The author acknowledges support from the Ministry of Research, Innovation and Digitalization, under the Romanian National Core Program LAPLAS VII -  Contract No. 30N/2023.

\end{document}